\documentclass[a4paper,11pt]{article}
\pdfoutput=1
\usepackage{jcappub}
\usepackage{aas_macros}
\usepackage{amsmath}
\usepackage{natbib}
\usepackage{hyperref}
\usepackage{amssymb}
\usepackage{epsfig}
\usepackage{graphicx}
\usepackage{subfigure}
\usepackage{afterpage}
\usepackage[T1]{fontenc}
\newcommand\Tstrut{\rule{0pt}{2.6ex}}

\title{\boldmath Cosmological Constraints on Horndeski Gravity in Light of GW170817}

\author[a,b,1]{C. D. Kreisch\note{Corresponding author.}}
\author[b,c]{and E. Komatsu}
\affiliation[a]{Princeton University,\\ 4 Ivy Lane, Princeton, NJ 08544 USA}
\affiliation[b]{Max-Planck-Institut f{\"u}r Astrophysik,\\ Karl-Schwarzschild Strasse 1, 85748 Garching, Germany}
\affiliation[c]{Kavli Institute for the Physics and Mathematics of the Universe (Kavli IPMU, WPI), Todai Institutes for Advanced Study, University of Tokyo, \\ Kashiwa 277-8583, Japan}
\emailAdd{ckreisch@astro.princeton.edu}
\emailAdd{komatsu@mpa-garching.mpg.de}

\abstract{The discovery of the electromagnetic counterpart to GW170817 severely constrains the tensor mode propagation speed, eliminating a large model space of Horndeski theory. We use the cosmic microwave background data from Planck and the joint analysis of the BICEP2/Keck Array and Planck, galaxy clustering data from the SDSS LRG survey, BOSS baryon acoustic oscillation data, and redshift space distortion measurements to place constraints on the remaining Horndeski parameters. We evolve the Horndeski parameters as power laws with both the amplitude and power law index free. We find a $95\%$~CL upper bound on the present-day coefficient of the Hubble friction term in the cosmological propagation of gravitational waves is $2.38$, whereas General Relativity gives $2$ at all times. While an enhanced friction suppresses the amplitude of the reionization bump of the primordial B-mode power spectrum at $\ell<10$, our result limits the suppression to be less than $0.8\%$. This constraint is primarily due to the scalar integrated Sachs-Wolfe effect in temperature fluctuations at low multipoles.}

\begin{document}
\maketitle
\flushbottom


\section{Introduction}
\label{sec:introduction}

The detection of gravitational waves opens a new window into constraining gravity. In general relativity (GR), the line element for scalar mode perturbations in the Newtonian gauge is given by $ds^2 = a^2\left(\tau\right)\left[-\left(1+2\Psi\right)d\tau^2+\left(1-2\Phi\right)dx^2\right]$. The line element for tensor mode perturbations is $ds^2 = a^2\left(\tau\right)\left[-d\tau^2+\left(\delta_{ij}+h_{ij}\right)dx^idx^j\right]$, where $h_{ij}=\pm h_{+},h_{\times}$ are small perturbations and are the divergenceless, traceless component of the metric. The linearized Einstein equation without a source takes the form of the wave equation $\Box h_{ij}=0$, where $\Box$ is the D'Alembertian. Then, going to Fourier space, the gravitational waves can be described by $\ddot{h}_{ij} + 2\frac{\dot{a}}{a}\dot{h}_{ij}+k^2h_{ij}=0$, where $k$ is the wavenumber. Dots throughout the paper denote derivatives with respect to conformal time.

Changing the gravitational theory modifies the propagation of gravitational waves. In Horndeski theory, the most general tensor-scalar theory in which the equations of motion are second order \citep{horndeski}, the tensor mode propagation equation becomes
\begin{eqnarray}
\label{eq:wave1}
\ddot{h}_{ij} + \left[2+\alpha_{\rm M}\left(a\right)\right]\mathcal{H}\dot{h}_{ij}+c_{\rm T}^2\left(a\right)k^2h_{ij}=0,
\end{eqnarray}
using the parameterization developed by \citep{bellini} and where $\mathcal{H}=\dot{a}/a$. The Planck mass run rate $\alpha_{\rm M}$ describes how the Planck mass evolves over time and contributes to the gravitational wave friction. It is defined as
\begin{align}
\label{eq:friction}
\alpha_{\rm M}=\frac{d\ln \left(M_*^2\right)}{d\ln a},
\end{align}
where $M_*^2$ is the effective Planck mass. If not constant, $\alpha_{\rm M}$ creates anisotropic stress in the Jordan frame.

The tensor speed~$c_{\rm T}$~is given by $c_{\rm T}^2=1+\alpha_{\rm T}$, where $\alpha_{\rm T}$ is the tensor speed excess that quantifies how much the gravitational wave speed deviates from that of light. Recent observations of GW170817 and its electromagnetic counterpart have placed the stringent bounds
\begin{align}
  -6 \times 10^{-15} \leq \alpha^{\rm T}_0 \leq 1.4 \times 10^{-15}.
\end{align}
This new constraint effectively eliminates all Horndeski theories with $\alpha^{\rm T}_0 \neq 0$ and alters the physically allowed values of the other Horndeski parameters \citep{alphaT}. See e.g. \cite{zbeyond} for an example of a non-trivial theory compatible with $c_\mathrm{T} = c$, \cite{lombriser1,lombriser2} for discussions of how gravitational wave detections impact Horndeski theory, \cite{DE1,DE2,baker2017} for good theoretical discussions about the impact of GW170817, and \cite{horndeskiconstraints} for observational constraints on Horndeski theory using only the simultaneous detection of GW170817 and GRB170817A. In this paper we use measurements of the cosmic microwave background (CMB), galaxy clustering, baryon acoustic oscillations (BAO), and redshift space distortions (RSD) to constrain the remaining Horndeski parameters in light of this discovery.

The remaining viable tensor mode propagation equation is
\begin{align}
\label{eq:wave}
\ddot{h}_{ij} + \left[2+\alpha_{\rm M}\left(a\right)\right]\mathcal{H}\dot{h}_{ij}+k^2h_{ij}=0.
\end{align}
Two additional parameters, the kineticity $\alpha_{\rm K}$ and braiding $\alpha_{\rm B}$, complete the parameterized Horndeski theory. Kineticity describes the scalar perturbations' kinetic energy. Large values reduce the scalar sound speed. The kinetic braiding parameter describes how the scalar and metric kinetic terms mix \citep[see e.g.][for good discussions]{braiding,bellini2015,bettoni}. A nonzero value indicates the clustering of dark energy. All four parameters are independent of each other and the background, which we choose to be $\Lambda\mbox{CDM}$. $\Lambda$CDM~GR cosmology is regained when all~$\alpha_i=0$.

In Section \ref{sec:data} we discuss our parameterization of the Horndeski parameters and what data we use to constrain them{, and in Section \ref{sec:stability} we define the model's stability constraints}. We present our results in Section \ref{sec:results} and discuss the source of constraining power in Section \ref{sec:constrain}. In Section \ref{sec:discussion} we compare our models and discuss our results. We summarize in Section \ref{sec:summary}. In Appendix \ref{sec:app_bg} we provide additional background information about Horndeski's theory and its implementation. Appendix \ref{sec:stability_constraints} details how kineticity affects the model's stability constraints and how this influences our results, and in Appendix \ref{sec:evolution} we discuss our evolution of the Horndeski parameters.

\section{Modeling \& Data Sets}
\label{sec:data}

We use {\fontfamily{pcr}\selectfont EFTCAMB}\footnote{\url{http://www.eftcamb.org/codes/download.html}, version 2.0} \citep{EFTCAMB1,EFTCAMB2}, a modified version of the Boltzmann code {\fontfamily{pcr}\selectfont CAMB} \citep{CAMB} that does not use a quasi-static approximation, as well as the complimentary {\fontfamily{pcr}\selectfont CosmoMC} version \citep{cosmomc}, to study the effects of modified gravity on perturbations. See Appendix \ref{sec:app_bg} for further detail on how the parameters are handled in the code. We set $R-1 \lesssim 0.03$, where $R$ is the Gelman-Rubin diagnostic \citep{gelman}, as our criterion to obtain convergence in the chains.

To explore the time evolution of the remaining Horndeski parameters and achieve a sufficiently large stable parameter space to perform an MCMC analysis (see Appendix \ref{sec:stability_constraints}), we evolve the Horndeski parameters as power laws:
\begin{align}
\alpha_{\rm K} & =\alpha^{\rm K}_0 a^{\kappa} \\
\alpha_{\rm B} & =\alpha^{\rm B}_0 a^{\xi} \\
\widetilde{M} & = \widetilde{M}_0 a^{\beta} \implies \alpha_{\rm M} =\frac{d\ln \left(M_*^2\right)}{d\ln a} = \widetilde{M}_0 a^{\beta} \frac{\beta}{1+\widetilde{M}_0 a^{\beta}},
\end{align}
where $M^2_*/m_0^2 = 1 + \widetilde{M}$ in the context of {\fontfamily{pcr}\selectfont EFTCAMB}, and $m_0^2$ is the Planck mass. $\widetilde{M}_0$ is the fractional deviation of $M_*^2$ from $m_0^2$ today. $\alpha_0^i$ denotes the parameter value today, and $a$ is the scale factor. This parameterization is similar to that used by the Planck collaboration for the Horndeski parameters \citep{planckMG}. While we do not know the exact functional form of the $\alpha_i$'s, Ref. \cite{linderchallenges} discusses the challenges of parameterizing them as a function of $\Omega_{\rm DE}$ and Ref. \cite{gleyzes2017} finds that evolving the $\alpha_i$'s as power laws, with both the amplitude and power law index free, is preferred over evolving them with simpler or more complex models.

The authors of \cite{bellini2015} parameterized the Horndeski parameters as a function of the dark energy density and found that $\alpha_{\rm K}$ could not be well constrained \citep[see also][]{bellinierr}. To limit the number of additional degrees of freedom in our analysis, we fix the evolution of $\alpha_{\rm K}$ with $\alpha^{\rm K}_0 = 0.1$ and $\kappa = 3$. We also choose to explore the case of $\alpha_0^{\rm B} = 0$ to probe a theory in which the primary modification is due to gravitational waves. Fixing $\alpha_0^{\rm B} = 0$ yields a perfect-fluid model that includes anisotropic stress \citep{bellini}. If both $\alpha_0^{\rm K} = 0$ and $\alpha_0^{\rm B} = 0$, the scalar propagation speed diverges. For this reason we fix the kineticity at a nonzero value throughout our analysis. See Appendix \ref{sec:stability_constraints} for a discussion of how kineticity affects the stable parameter space given the imposed stability conditions defined in Section \ref{sec:stability}. We vary the standard cosmological parameters $\Omega_b h^2$, $\Omega_c h^2$, $\theta$, $\tau$, ${\rm{ln}}(10^{10} A_s)$, $n_s$, as well as the tensor-to-scalar ratio $r$, Planck calibration $y_{\rm cal}$, the dust power ($\ell = 80,\,\nu=353\,\mathrm{GHz}$) $A_{B,\mathrm{dust}}$, and the dust frequency scaling parameter $\beta_{B,\mathrm{dust}}$. We choose not to vary $w_0$ and $w_a$ to minimize the number of free parameters in our analysis and to focus on the propagation of gravitational waves. In \autoref{tab:priors} we list our adopted prior cutoffs for the different parameters. All parameters have uniform priors except for $y_{\rm cal}$ and $\beta_{B,\mathrm{dust}}$, which have the Gaussian priors $y_{\rm cal} = 1.0000 \pm 0.0025$ and $\beta_{B,\mathrm{dust}} = 1.59 \pm 0.11$, respectively.

\begin{table}[tbp]
\caption{Adopted MCMC Priors \label{tab:priors}}
\center
\begin{tabular}{cc}
\hline
 Parameter & Prior \\
\hline \\[-1.5ex]
{$\widetilde{M_0}    $} & $\left[-1,5 \right]$ \\
{$\beta$} & $\left[0,20 \right]$ \\
$\alpha_{0}^{\rm B}$ & $\left[-7,2 \right]$ \\
$\xi$ & $\left[0,20 \right]$ \\[+1.5ex]
\hline \\[-1.5ex]
{$\Omega_b h^2$} & $\left[0.005, 0.100\right]$ \\
{$\Omega_c h^2$} & $\left[0.001, 0.990\right]$ \\
{$100\theta_{MC}$} & $\left[0.5, 10\right]$ \\
{$\tau$} & $\left[0.01, 0.80\right]$ \\
{${\rm{ln}}(10^{10} A_s)$} & $\left[2, 4\right]$ \\
{$n_s$} & $\left[0.8, 1.2 \right]$ \\
{$r$} & $\left[0, 1 \right]$ \\[+1.5ex]
\hline \\[-1.5ex]
{$y_{\rm cal}$} & $\left[0.9, 1.1 \right]$ \\
{$A_{B,\mathrm{dust}}$} & $\left[0, 15 \right]$ \\
{$\beta_{B,\mathrm{dust}}$} & $\left[1.04, 2.14 \right]$ \\
\hline
\end{tabular}
\end{table}

Given the direct effect the Horndeski parameters have on tensor perturbations \citep[see \autoref{eq:wave} and e.g.][]{amendola,raverispeed,pettorino}, we include in our analysis the B-mode data from the joint analysis of the BICEP2/Keck Array and Planck \citep{BKP}. Because the Horndeski parameters influence the scalar perturbations, as well, we use the 2015 Planck low-$\ell$ CMB temperature and polarization data, high-$\ell$ temperature data, and the lensing potential measurements \citep{planckdata,plancklens}.

It is reasonable to posit the CMB places optimal constraints on the remaining Horndeski parameters at the epoch of decoupling. We studied the optimal pivot scale to measure $\alpha_{\rm B}$ and $\alpha_{\rm M}$ and find that they are uncorrelated at present and are best constrained at late times rather than during recombination. See Appendix \ref{sec:evolution} for a further discussion. The constraints primarily come from the late integrated Sachs-Wolfe (ISW) effect (see Section \ref{sec:constrain}). The pivot scale is model dependent, however, and should be re-examined when the evolution of the Horndeski parameters is better determined. To take advantage of this late time sensitivity, we include the SDSS LRG DR4 \citep[mpk,][]{mpkdata} and BOSS BAO and RSD data sets in our MCMC analysis \citep{BAO2,BAO1,RSD1,RSD2}.

\section{Stability Constraints}
\label{sec:stability}

Several viability priors can be set by {\fontfamily{pcr}\selectfont EFTCAMB} to ensure the parameter space yields a stable theory. We enforce the following constraints \citep[see][for a full description]{EFTCAMB1,eftcambnotes,bellini}:
\begin{enumerate}
\item Physical stability: the theory must have a positive effective Newton's constant (i.e., $1+\Omega>0$), and avoid ghost and gradient instabilities. Ghost instabilities arise when the kinetic energy becomes negative, and gradient instabilities occur when the squared speed of sound for perturbations becomes negative \cite[see \S 3.3 of][]{bellini,EFTCAMB1}.
\item Mathematical stability: neither the coefficient of $\ddot h_{ij}$ nor the coefficient of $\ddot\pi$ may equal $0$, ensuring the tensor perturbation and $\pi$ field equations are well defined, respectively \cite{EFTCAMB1,eftcambnotes,stab_math}. In this work we exclude all exponential growth of the $\pi$ field perturbations, {including those due to tachyon instabilities\footnote{{Some argue that tachyon instabilities are not harmful because they produce perturbation growth on large scales which is, thus, bounded \citep[see e.g.][]{HICLASS}. In contrast are gradient instabilities that produce unbounded perturbation growth on small scales. Nevertheless, in this paper we follow the framework of EFTCAMB, in which the mathematical stability conditions include the tachyon stability requirement.}}}. For details of the mathematical stability conditions, see Equations 40 and 52 in \cite{eftcambnotes}, as well as the corresponding $\pi$ field equation discussion in section IV D and viability condition discussion in section IV F of \cite{eftcambnotes}.
\item We require a positive matter density and dark energy density with $\omega_{\rm DE}\leq -1/3$ at all times.
\end{enumerate}

We do not restrict our analysis to regions of parameter space where $c_s^2\le 1$ or $m_\pi^2\ge 0$. Subluminal sound speeds are required for the theory to be UV complete through standard methods \citep{uv,bellini}. The authors of \cite{superlum} have shown, however, that Horndeski theories will always have regions of parameter space that include superluminal sound speeds \citep[see also][for a brief discussion]{bellini}. If the scalar field couples to matter, its speed and mass are more complicated to compute than if the field were in a vacuum. They cannot be directly read from the $\pi$ propagation equation because the scalar degree of freedom of the theory is a combination of $\pi$ and the matter fields. Enforcing $c_s^2\le 1$ in {\fontfamily{pcr}\selectfont EFTCAMB} would not put a limit on the true scalar sound speed in the non-minimal coupling scenario \citep{speed}. We note that restrictions on $c_s$ can have a severe impact on the stable parameter space. The authors of \cite{salvatelli2016} have shown that restricting the scalar field to propagate subluminally yields a stable parameter space that is a very small subsection of the parameter space allowed when the scalar field is free to have any real sound speed.

\section{Results}
\label{sec:results}

\begin{figure*}[t]
\begin{center}
\includegraphics[width=\textwidth]{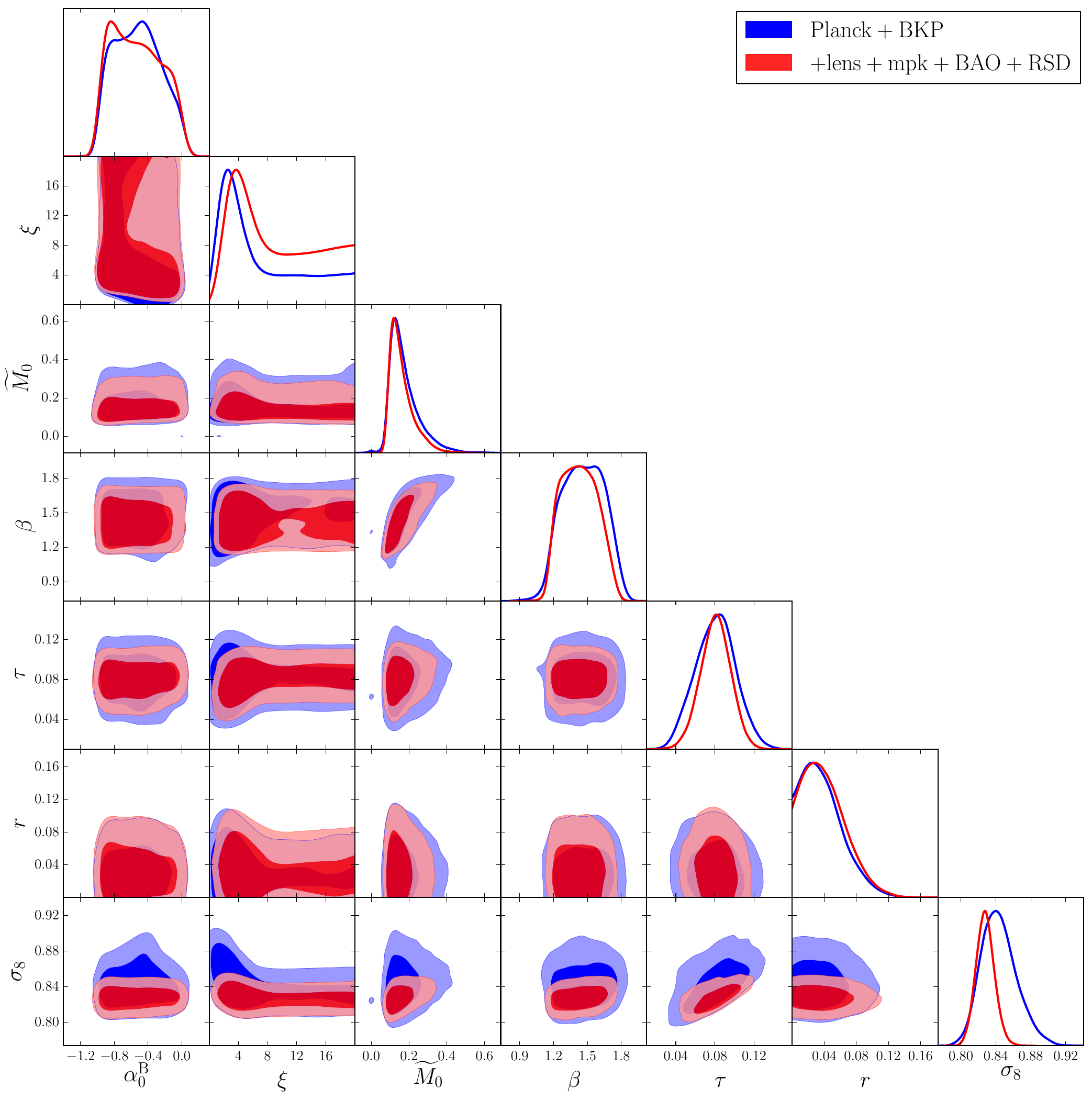}
\caption{Constraints on the Horndeski parameters, $\tau$, $r$, and $\sigma_8$ for $\alpha_0^{\mathrm K} = 0.1$ and $\alpha_0^{\mathrm B} \neq 0$.}
\label{fig:contours}
\end{center}
\end{figure*}

\begin{figure}[t]
\begin{center}
\includegraphics[width=0.5\textwidth]{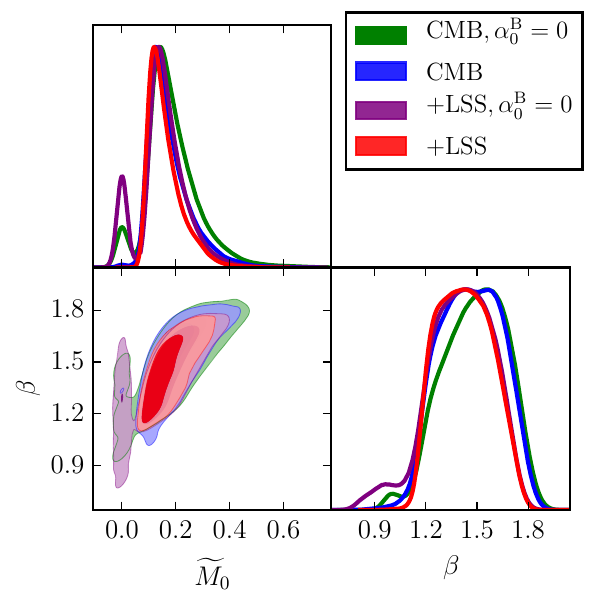}
\caption{Constraints on the Horndeski friction parameters for $\alpha_0^{\mathrm K} = 0.1$ and both braiding cases. The bimodality in the $\widetilde{M}_0$ posterior is due to stability constraints. We note that smoothing for plotting purposes reduces the amplitude of the leftmost mode.}
\label{fig:contours_noB}
\end{center}
\end{figure}

In \autoref{fig:contours} and \autoref{fig:contours_noB} we show the resulting posterior probability distributions for the cases with $\alpha_0^{\rm B} \neq 0$ and $\alpha_0^{\rm B} = 0$, respectively. CMB denotes the low-$\ell$ TEB, high-$\ell$ TT \citep{planckdata}, and BKP \citep{BKP} data set combination. LSS denotes combining the lensing \citep{plancklens}, mpk \citep{mpkdata}, BAO \citep{BAO2,BAO1}, and RSD \citep{RSD1,RSD2} data sets. Constraints on the friction and braiding parameters are quoted in \autoref{tab:params} and \autoref{tab:params_noB}. See \cite{cluster} for constraints on $\alpha_{\rm B}$ from galaxy cluster observations in light of the observation of the electromagnetic counterpart to GW170817.

One may be surprised to see that the bounds on $\widetilde{M}_0$ and $\alpha_0^{\mathrm{M}}$ exclude their GR values at $95\%$ confidence in the case of $\alpha_0^{\mathrm{B}} \neq 0$. However, these lower bounds are driven largely by stability constraints on the model when $\alpha_0^{\rm K}=0.1$. For this kineticity, as $\widetilde{M}_0 \to 0^+$, $\alpha_0^{\rm B} \to 0^+$, as well. Thus, for the $\alpha_0^{\rm B} \neq 0$ case, there is not a large enough stable region around $\widetilde{M}_0=0$ for the posterior to show a nonzero probability.

$\alpha_0^{\rm M}$ has an upper limit of $0.38$ and $0.41$ at $95\%$ confidence for $\alpha_0^{\mathrm{B}} \neq 0$ and $\alpha_0^{\mathrm{B}} = 0$, respectively, for the CMB+LSS data sets. The stable parameter space for $\widetilde{M}_0$ is large (see Appendix \ref{sec:stability_constraints}), so the upper $95\%$ CL constraints for this parameter are driven by data. Matter clustering and the late ISW effect are both sensitive probes of $\widetilde{M}_0$ (see Section \ref{sec:constrain}).

For $\alpha_0^{\mathrm{B}} = 0$, $\widetilde{M}_0$ and $\alpha_0^{\mathrm{M}}$ are consistent with GR. See also the bimodal posterior for $\widetilde{M}_0$ in \autoref{fig:contours_noB} when $\alpha_0^{\mathrm{B}} = 0$. When $\alpha_0^{\rm B} = 0$, $\widetilde{M}_0$ is stable at $0$ and values near $0$, allowing for the bimodal posterior in \autoref{fig:contours_noB}. The posterior is bimodel rather than smooth because there is a small number of stable parameter values when $0 \lesssim \widetilde{M}_0 \lesssim 0.1$. Indeed, the $\widetilde{M}_0$ posterior for the $\alpha_0^{\rm B} \neq 0$ case and the rightmost mode in the posterior for the $\alpha_0^{\rm B} = 0$ case rise near $\widetilde{M}_0 = 0.1$.

The shape of this stable parameter region near $\widetilde{M}_0=0$ is purely driven from the choice of $\alpha_0^{\rm K}$. When $\alpha_0^{\rm K}=0$, values between $\widetilde{M}_0=0$ and $\widetilde{M}_0 \approx 0.1$ are stable, eliminating the bottleneck effect in the $\widetilde{M}_0-\alpha_0^{\rm B}$ stable space as $\widetilde{M}_0 \to 0^+$. However, to investigate the case of $\alpha_0^{\rm B}=0$, $\alpha_0^{\rm K}$ must be nonzero since $\alpha_0^{\rm B}=0$ and $\alpha_0^{\rm K}=0$ simultaneously are unstable. See Appendix \ref{sec:stability_constraints} for a further discussion.

Both $\xi$, the power law index for the braiding parameter, and $\beta$, the power law index related to friction, are relatively unconstrained by the data. In fact, $\xi$ cannot be constrained at the $95\%$ CL for the $\alpha_0^{\rm B} \neq 0$ case with CMB data alone. At $68\%$ confidence, $\xi > 2.82$ for this case. Note that the nonzero probability for $\xi=0$ in \autoref{fig:contours} is due to smoothing artifacts. The constraints on $\beta$ are almost purely from stability constraints (see Appendix \ref{sec:stability_constraints}). There is a small stable region for $\beta$ near $0$ if $\xi$ is near $0$, but the data did not prefer this region. The $\beta$ posterior bounds in \autoref{fig:contours} and \autoref{fig:contours_noB} correspond to the boundaries of the remaining portion of the stable parameter space. $\xi$ has a large stable parameter space, so the best-fit appears driven by the data. However, the plateau in the posterior for large $\xi$ is due to the data being unable to further constrain the parameter.

Larger values of $\beta$ are preferred for larger $\widetilde{M}_0$. This suggests that, for a power law evolution, the data prefer to minimize the deviation from $m_0^2$ at early times. The $\xi-\alpha_0^{\mathrm B}$ contour in \autoref{fig:contours} shows a similar trend for $\xi \lesssim 2$ as $\alpha_0^{\mathrm B}$ deviates farther from $0$.

\begin{table}[tbp]
\caption{Parameter $95\%$ Confidence Limits, $\alpha_0^{\rm B} \neq 0$ \label{tab:params}}
\center
\begin{tabular}{ccc}
\hline
 Parameter & Planck + BKP & + lens + mpk + BAO + RSD \\
\hline \\[-1.5ex]
{$\widetilde{M_0}    $} & $0.17^{+0.18}_{-0.11}$ & $0.160^{+0.14}_{-0.091}$ \\
{$\beta$} & $1.46^{+0.32}_{-0.31}$ & $1.43^{+0.28}_{-0.27}$ \\
$\alpha_{0}^{\rm M}$ & $0.22^{+0.23}_{-0.16}$ & $0.20^{+0.18}_{-0.12}      $ \\[+1.5ex]
\hline \\[-1.5ex]
{$\alpha_0^{\rm B}$} & $-0.53^{+0.50}_{-0.46}$ & $-0.54^{+0.52}_{-0.46}$ \\
{$\xi$} & $-$ & $> 2.16$ \\[+1.5ex]
\hline
\end{tabular}
\end{table}

\begin{table}{}
\caption{Parameter $95\%$ Confidence Limits, $\alpha_0^{\rm B} = 0$ \label{tab:params_noB}}
\center
\begin{tabular}{ccc}
\hline
 Parameter & Planck + BKP & + lens + mpk + BAO + RSD \\
\hline\\[-1.5ex]
{$\widetilde{M_0}    $} & $0.18^{+0.21}_{-0.21}$ & $0.14^{+0.18}_{-0.18}      $ \\
{$\beta$} & $1.49^{+0.33}_{-0.34}$ & $1.40^{+0.36}_{-0.45}$ \\
$\alpha_{0}^{\rm M}$ & $0.23^{+0.25}_{-0.26}$ & $0.18^{+0.23}_{-0.22}$ \\[+1.5ex]
\hline
\end{tabular}
\end{table}

\section{Sources of Constraints}
\label{sec:constrain}

\begin{figure}[t]
\begin{center}
\includegraphics[width=0.7\textwidth]{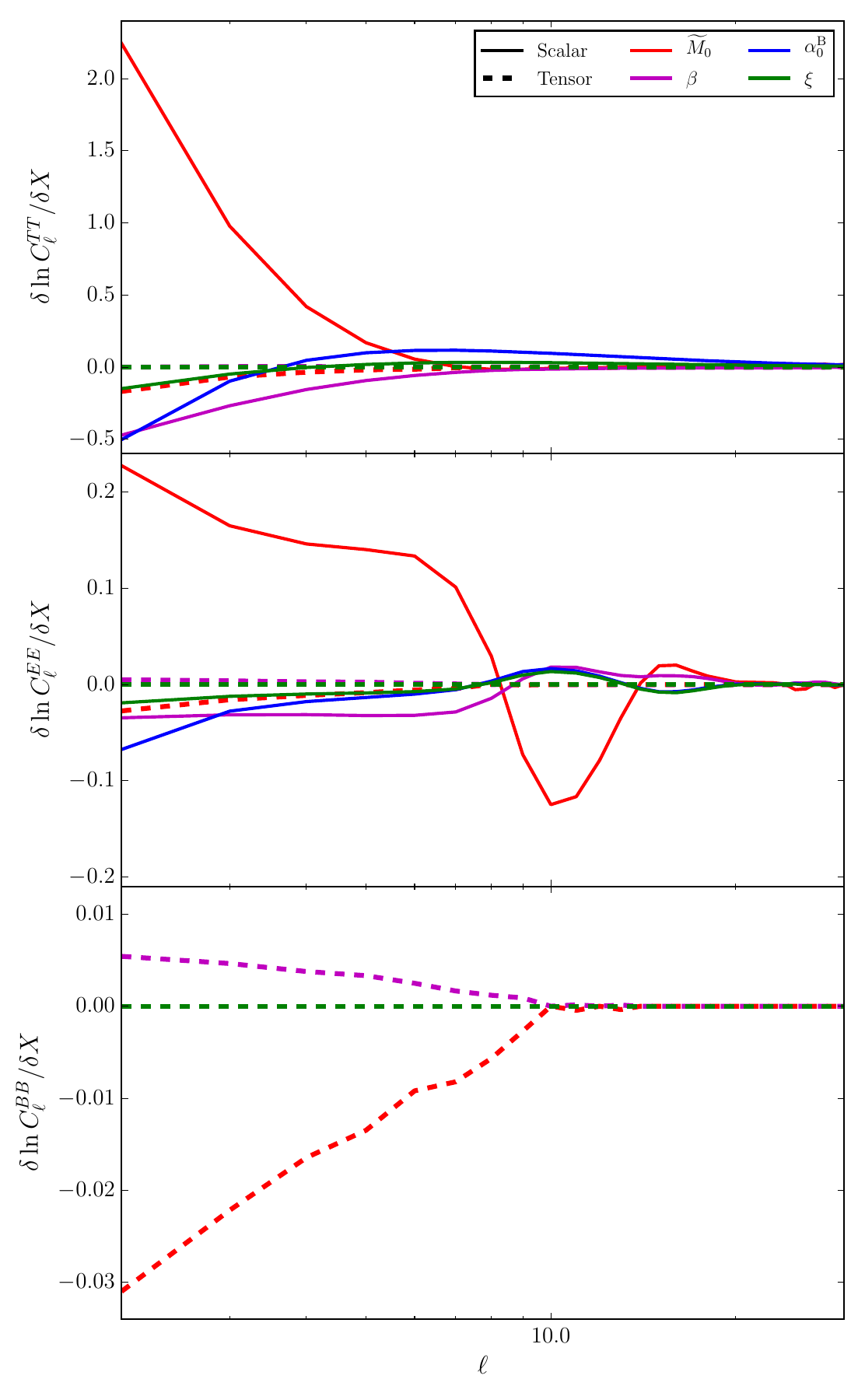}
\caption{Sensitivity of the CMB perturbations to the Horndeski parameters as a function of scale. Colors denote which Horndeski parameter was varied while all other parameters were held constant. Solid (dashed) lines correspond to scalar (tensor) modes. Note that what is shown is equivalent to the derivative of a fractional deviation, i.e. $\frac{1}{C_\ell}\delta C_\ell / \delta X$.}
\label{fig:scal_tens_CMB}
\end{center}
\end{figure}

Where do the constraints on the Horndeski parameters come from? To identify the source of dominating constraining power, we compute power spectra derivatives at our fiducial cosmology for the temperature, E-mode polarization, B-mode polarization, lensing potential, and matter power spectra in Figures \ref{fig:scal_tens_CMB}-\ref{fig:dPdx}. Our fiducial values for the Horndeski parameters, tensor-to-scalar ratio, and tensor tilt are listed in \autoref{tab:fiduc}, and we use the best-fit Planck TT+lowP+lensing+ext $\Lambda\mbox{CDM}$ parameter values \citep{planckparam}. The fiducial Horndeski values were chosen based on the posteriors in \autoref{fig:contours} and \autoref{fig:contours_noB}. Computing the relative difference between the Horndeski and fiducial (rather than GR) spectra ensures all the spectra have a stable theory while only varying a single parameter.

The power spectra are most affected at large scales by the Horndeski parameters, and the CMB scalar modes and large scale structure (LSS) are significantly more sensitive to these parameters than the CMB tensor modes.

\begin{table}[h]
\caption{Fiducial Parameters \label{tab:fiduc}}
\center
\begin{tabular}{l @{\qquad} c} \\[-5.0ex]
\hline
 Parameter & Value \\
\hline
$\alpha_0^K$  & $0.1$ \Tstrut\\
$\kappa$      & $3.0$ \\
$\alpha_0^B$ & $-0.5$ \\
$\xi$  & $3.0$ \\
$\widetilde{M}_0$  & $0.15$ \\
$\beta$ & $1.5$ \\
\hline\\[-1.5ex]
$r$  & $0.05$ \\
$n_t$ & $0$ \\
\hline
\end{tabular}
\end{table}

\subsection{Cosmic Microwave Background}
\subsubsection{Scalar Perturbations}
In Figure \ref{fig:scal_tens_CMB} we show the sensitivity of the unlensed temperature and polarization anisotropies to changes in the Horndeski parameters in the neighborhood of the fiducial parameters, which is a good proxy for the constraining power near the best-fit values. In general, the CMB anisotropies are most sensitive to the Horndeski parameters at $\ell \lesssim 10$, though measurements at such scales are limited by cosmic variance. The perturbations are most sensitive to the friction parameter $\widetilde{M}_0$, and $\widetilde{M}_0$ has a more dramatic effect on the scalar temperature anisotropies than on polarization, lensing, and matter clustering.

\begin{figure}[t!]
\begin{center}
\includegraphics[width=0.673\textwidth]{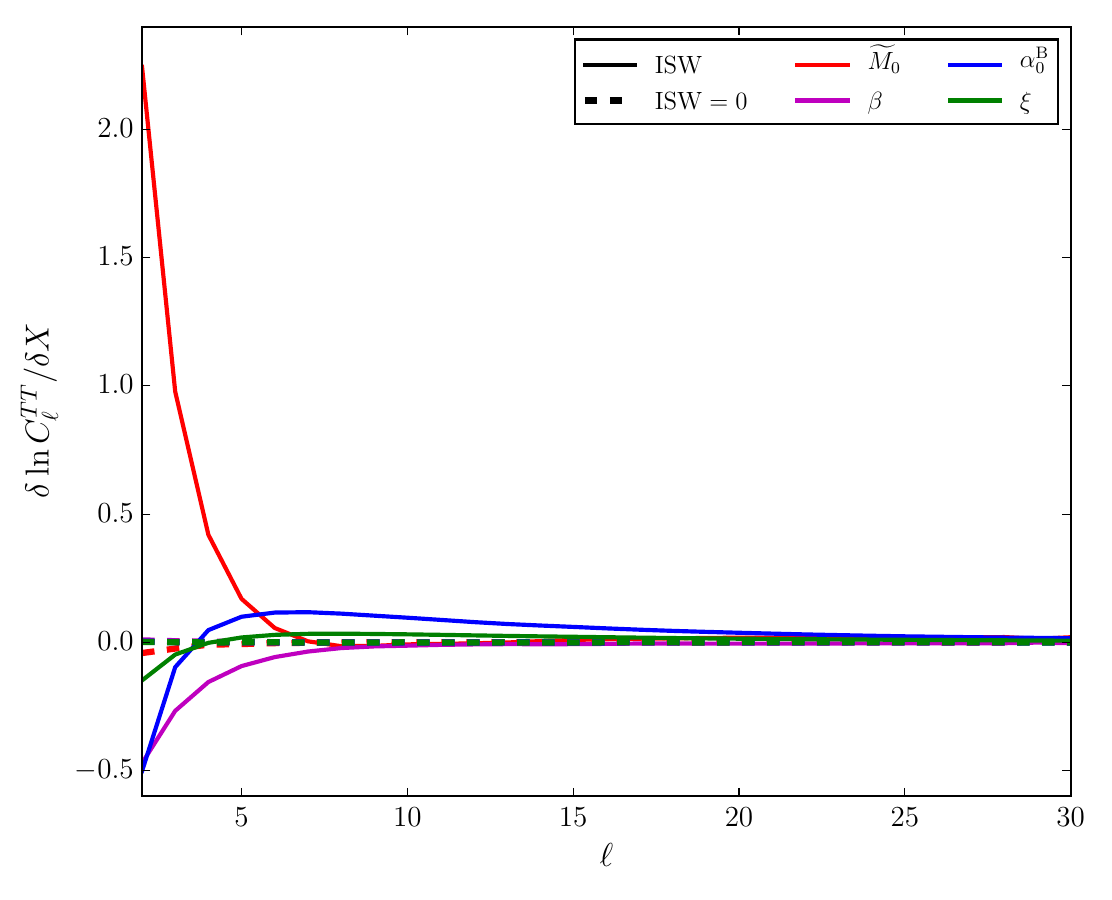}
\caption{Sensitivity of the ISW effect to the Horndeski parameters as a function of scale. Colors denote which Horndeski parameter was varied while all other parameters were held constant. Solid (dashed) lines correspond to including (excluding) the ISW. Removing the ISW from the TT power spectrum eliminates the sensitivity of the temperature anisotropies to the Horndeski parameters.}
\label{fig:noISW}
\end{center}
\end{figure}

The sensitivity of the scalar TT power spectrum to the Horndeski parameters at low-$\ell$ suggests the ISW effect may be the source of sensitivity. In Figure \ref{fig:noISW} we show power spectra derivatives with and without the ISW effect. Removing the ISW effect erases any sensitivity the scalar TT power spectrum had to the Horndeski parameters, including braiding, indicating the ISW effect is the primary source of constraint for the temperature anistropies. Because a nonzero Planck-mass run rate creates anisotropic stress, the evolution of the Bardeen potentials changes over time \citep{renk, HICLASS}. The anistropic stress constraint from the spatial traceless component of the Einstein equations makes this clear \citep{bellini}:
\begin{align}
  \Psi-\Phi-\alpha_{\rm M} H v_{\rm X} = \tilde{p}_{\rm m} \pi_{\rm m},
\end{align}
where $v_{\rm X}=-\frac{a\delta \phi}{\dot{\phi}}$ is the scalar velocity potential, and $H = \mathcal{H}/a$. Friction alters the relationship between the Bardeen potentials, leading to a change in the ISW and, thus, the temperature anisotropies. As seen in \autoref{fig:noISW}, increasing braiding decreases the ISW effect, reducing the power of the temperature anisotropies at low-$\ell$ \citep[see][for a further discussion of $\alpha_{\rm M}$, $\alpha_{\rm B}$, and the ISW effect]{renk}.

Friction's effect on the temperature anisotropies via the ISW also changes the local temperature quadrupole seen by electrons during reionization, altering the low-$\ell$ E-mode. Scattering of the reionization electrons off the quadrupole produces additional polarization at the scale that enters the horizon during reionization. Increasing $\widetilde{M}_0$ makes the reionization bump peakier for the scalar E-modes, boosting the power for $\ell < 10$ and damping the power near $\ell = 10$.


\begin{figure}[t]
\begin{center}
\includegraphics[width=0.673\textwidth]{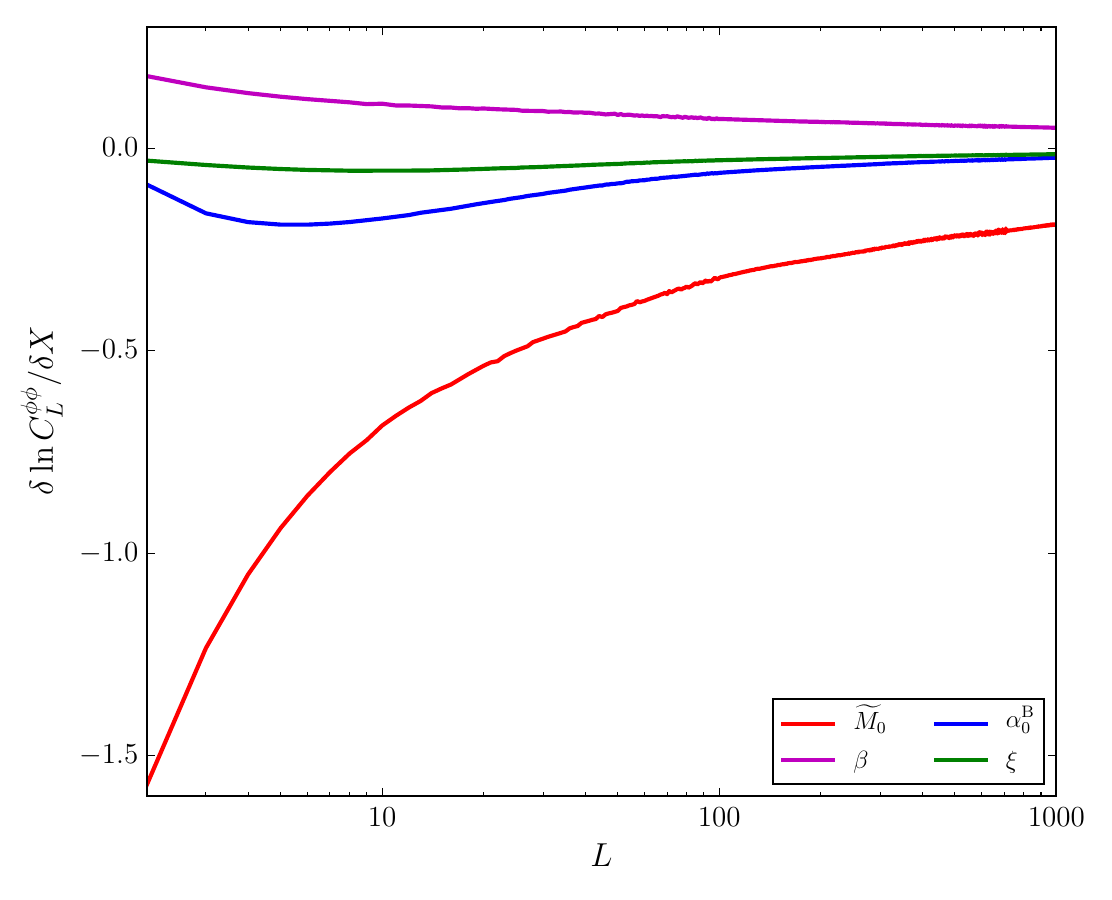}
\caption{Sensitivity of the lensing potential to the Horndeski parameters. Colors denote which Horndeski parameter was varied while all other parameters were held constant.}
\label{fig:deltalens}
\end{center}
\end{figure}

\begin{figure}[h]
\begin{center}
\includegraphics[width=0.673\textwidth]{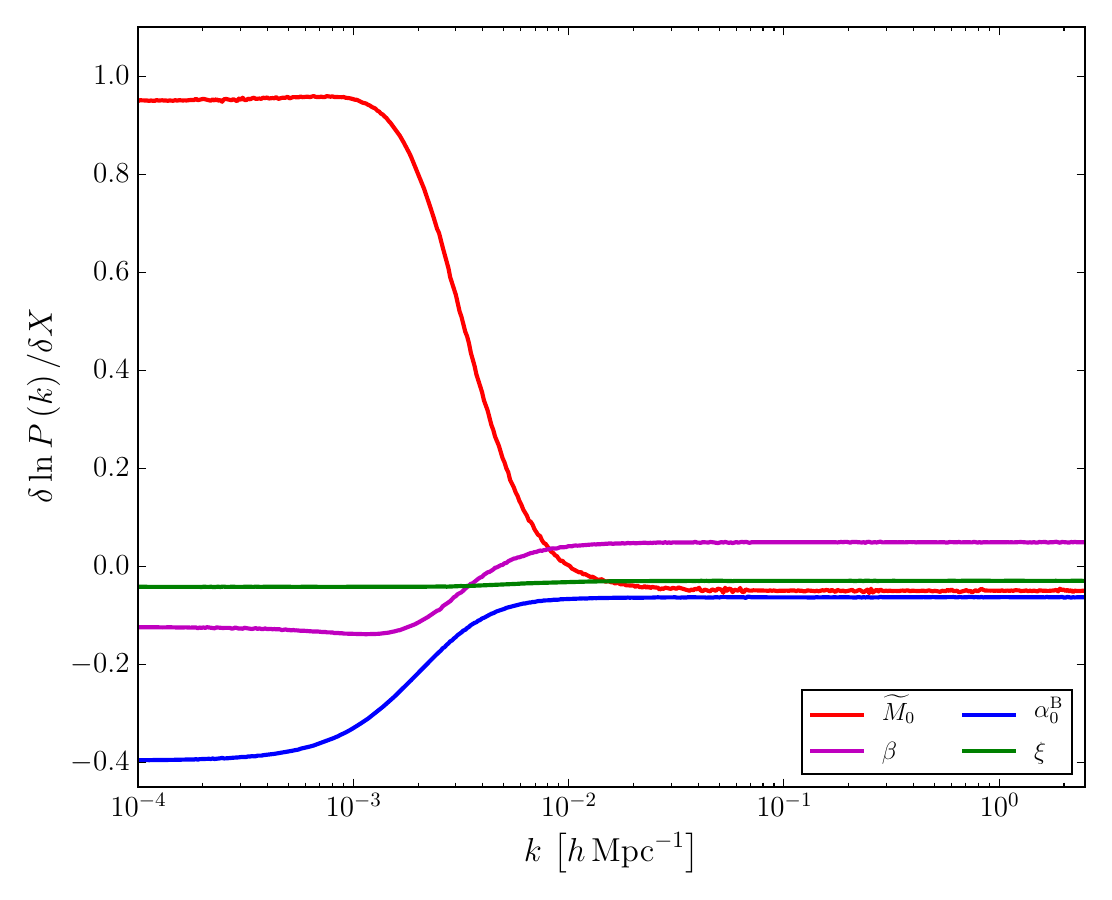}
\caption{Sensitivity of matter clustering to the Horndeski parameters. Colors denote which Horndeski parameter was varied while all other parameters were held constant.}
\label{fig:dPdx}
\end{center}
\end{figure}

\subsubsection{Tensor Perturbations}

Braiding only affects the scalar modes. Increasing the friction term $\widetilde{M}_0$ dampens the tensor perturbations, decreasing the B-mode and tensor E-mode amplitudes (see middle and bottom panels in \autoref{fig:scal_tens_CMB}). This is expected since the form of \autoref{eq:wave} is that of a damped driven oscillator \citep[see also][]{pettorino,tensorpert}. For the tensor mode polarization, increasing $\widetilde{M}_0$ decreases the reionization bump. See e.g. \cite{amendola} and \cite{pettorino} for a further discussion on how friction affects the reionization peak.

Only gravitational waves outside the horizon at recombination affect the temperature anisotropies since the gravitational waves decay and oscillate as soon as they enter the horizon. Since friction is a damping term for the gravitational waves, the tensor temperature power spectrum is damped for low-$\ell$, as well. Nonetheless, friction has a more dramatic effect on the scalar modes, so the scalar temperature and E-mode polarization dominate the CMB constraints.

\subsection{Large Scale Structure}

Increasing friction damps the lensing potential at all scales, but most significantly at large angular scales (see \autoref{fig:deltalens}). This occurs because $|\Phi+\Psi|$ is damped from increasing the friction. The lensing potential is almost as sensitive to the friction parameters as the scalar TT power spectrum at large scales, but present lensing potential measurements do not cover multipoles $L<40$ and so the lensing potential does not have strong constraining power.


Both $\alpha_{\rm B}$ and $\alpha_{\rm M}$ alter the growth rate \citep{bellini}. In the neighborhood of our fiducial cosmology,
increasing $\widetilde{M}_0$ boosts clustering for $k \lesssim 10^{-2}\,h\,\mathrm{Mpc}^{-1}$ (see \autoref{fig:dPdx}). Increasing $\alpha_0^{\rm B}$ so that it is closer to GR reduces clustering on similar scales. The matter power spectrum shows weak sensitivity to the Horndeski parameters for the scales directly probed by the LSS measurements. The dominating constraining power on the Horndeski parameters from these measurements comes through their constraints on $\sigma_8$.

\section{Discussion}
\label{sec:discussion}

\begin{table}{}
\caption{Model Comparison: Values of $\Delta \mathrm{AIC} = \mathrm{AIC}_{\alpha_{\rm B}\neq 0}-\mathrm{AIC}_{\alpha_{\rm B} = 0} = \Delta \chi^2 + 4$\label{tab:like_ratios}}
\center
\begin{tabular}{ccc}
\hline
 Observable & Planck + BKP & + lens + mpk + BAO + RSD \\
\hline
{${\mathrm{BKP}}$} & $3.95$ & $4.02$ \\
${\mathrm{high-}\ell\,\mathrm{TT}}$ & $2.63$ & $3.95$ \\
${\mathrm{low-}\ell\,\mathrm{TEB}}$ & $3.40$ & $3.75$ \\
{${\mathrm{lens}}$} & $-$ & $3.86$ \\
{${\mathrm{mpk}}$} & $-$ & $3.96$ \\
{${\mathrm{BAO}}$} & $-$ & $3.99$ \\
{${\mathrm{RSD}}$} & $-$ & $4.18$ \\
{${\mathrm{Total}}$} & $1.97$ & $3.72$ \\
\hline
\end{tabular}
\end{table}

To compare the two models we can use the Akaike information criterion (AIC) \citep{AIC}, defined as:
\begin{align}
  \mathrm{AIC}=-2{\rm ln}\left(\mathcal{L}\right)+2k=\chi^2+2k,
\end{align}
where $\chi^2 = \chi^2_\mathrm{BKP} + \chi^2_{\mathrm{high-}\ell\,\mathrm{TT}} + \chi^2_{\mathrm{low-}\ell\,\mathrm{TEB}}$ for the CMB data set combination and $\chi^2 = \chi^2_\mathrm{BKP} + \chi^2_{\mathrm{high-}\ell\,\mathrm{TT}} + \chi^2_{\mathrm{low-}\ell\,\mathrm{TEB}} + \chi^2_\mathrm{lens} + \chi^2_\mathrm{mpk} + \chi^2_\mathrm{BAO} + \chi^2_\mathrm{RSD}$ for the CMB+LSS data set combination. Then,
\begin{align}
  \Delta \mathrm{AIC} = \mathrm{AIC}_{\alpha_{\rm B}\neq 0}-\mathrm{AIC}_{\alpha_{\rm B} = 0} = \Delta \chi^2 + 4
\end{align}
where $\mathcal{L}$ is the maximum-likelihood and $k$ is the number of fit parameters, yielding $\Delta \mathrm{AIC} = 1.97$ for the CMB data sets and $\Delta \mathrm{AIC} = 3.72$ after including large scale structure measurements. The AIC takes into account how well the model fits the data while incorporating a penalty proportional to the number of parameters fit. When comparing two models, the lower AIC corresponds to the preferred model. In principle, the Bayes factor should be used to compare the models and the AIC proves only an approximation to it \citep[see, e.g., the introduction of][for the Bayes factor]{bayes}.

In \autoref{tab:like_ratios} we list the individual $\Delta \mathrm{AIC}$ values for each data set used in the CMB and CMB+LSS combinations, as well as the total $\Delta \mathrm{AIC}$ value for both combinations. All $\Delta \mathrm{AIC}$ values are positive, indicating the $\alpha_{\rm B} = 0$ model is preferred for all data sets. For this case the data are consistent with GR.

Incorporating the LSS data leads to a lower preferred value and upper bound on $\widetilde{M}_0$ (see \autoref{tab:params} and \autoref{tab:params_noB}). Increasing $\widetilde{M}_0$ boosts the power of the matter power spectrum for large scales, increasing $\sigma_8$ (see \autoref{fig:dPdx}). A weak correlation between $\widetilde{M}_0$ and $\sigma_8$ is visible in the lower boundary of the relevant contour in \autoref{fig:contours}. The RSD measurement's preference for a lower $\sigma_8$ helps shrink the contour and leads to a slightly lower preferred $\widetilde{M}_0$.

\begin{figure*}[t!]
\begin{center}
\includegraphics[width=\textwidth]{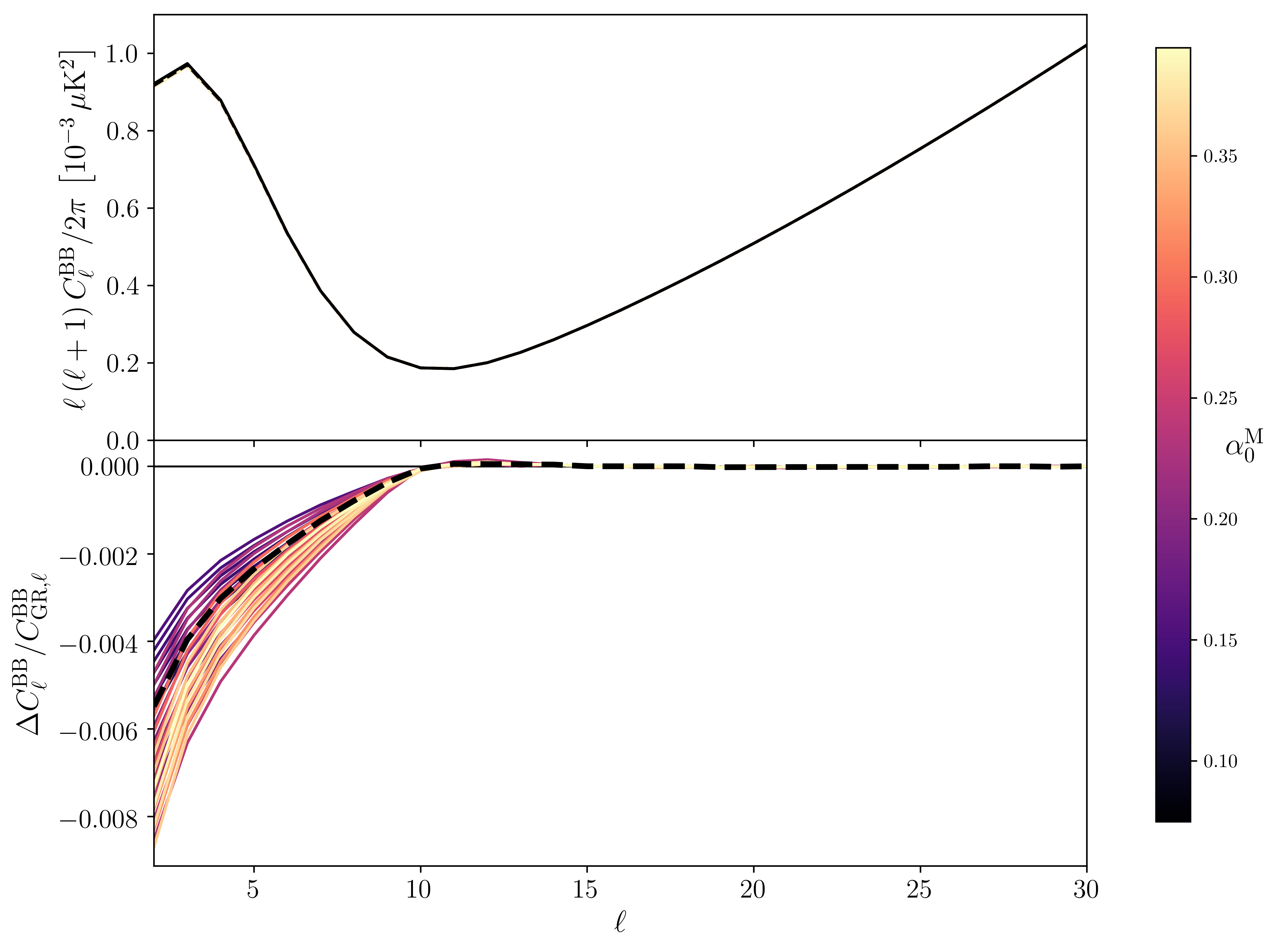}
\caption{Primordial B-mode spectra for different values of $\alpha_0^{\rm M}$ allowed at the $95\%$ CL by the CMB+LSS data sets for $\alpha_0^{\rm B} \neq 0$ with all other parameters fixed. The bottom panel shows the relative difference to GR. The dashed line corresponds to the best-fit values of $\widetilde{M}_0$ and $\beta$, the power law index related to friction, for the CMB+LSS data sets for $\alpha_0^{\rm B} \neq 0$.}
\label{fig:B_modes}
\end{center}
\end{figure*}

A primary goal of current and future CMB polarization experiments is to make the first detection of B-mode polarization from primordial gravitational waves. With our constraints we can investigate the impact of the Horndeski parameters on such experiments. In \autoref{fig:B_modes} we plot the primordial B-modes while varying $\alpha_0^{\rm M}$ in the $95\%$ CL range allowed by the CMB+LSS data set combination for $\alpha_0^{\rm B} \neq 0$ to show the range of B-mode spectra consistent with the data. We use $r=0.05$, $n_t=0$, the best-fit Planck TT+lowP+lensing+ext $\Lambda\mbox{CDM}$ parameter values \citep{planckparam}, and our best-fit $\alpha_0^{\rm B}$ and $\xi$, the power law index for braiding, values for the CMB+LSS data set combination. Even for the largest $\alpha_0^{\rm M}$ allowed by the data, the deviations from GR are less than $1\%$ at the reionization bump at $\ell<10$, which is smaller than the cosmic variance. Thus, even without a detection of primordial B-modes to date, we can conclude that the effect of the Horndeski parameters on the primordial B-mode polarization is constrained to be insignificant.

\section{Summary}
\label{sec:summary}

With a fixed kineticity of $\alpha_{\rm K} = 0.1a^3$ we have shown:
\begin{itemize}
  \item The friction $\alpha_0^{\rm M}$ has a $95\%$ CL upper limit of $0.38$ and $0.41$ when $\alpha_0^{\rm B} \neq 0$ and $\alpha_0^{\rm B} = 0$, respectively, when using both CMB and LSS data.
  \item The lower $95\%$ CL limit for $\alpha_0^{\rm M}$ excludes GR for the $\alpha_0^{\rm B} \neq 0$ case but not for the $\alpha_0^{\rm B} = 0$ case. We believe this is primarily due to {tachyon instabilities} (defined in Section \ref{sec:stability}) imposed by fixing $\alpha_0^{\rm K}=0.1$.
  \item The effects of Horndeski theory on primordial B-modes are constrained by CMB and LSS data to be insignificant with $95\%$ confidence.
\end{itemize}

It is important to remember that even when using Horndeski theory, making different assumptions about the $\alpha_i$ parameters can yield dramatically different results. See e.g. the tight $\alpha_0^{\rm M}$ constraint by \cite{planckMG}, the negative values allowed by \cite{pettorino}, \cite{bellini2015}, and \cite{bellinierr}, and the constraints found by \cite{salvatelli2016}. We stress that choice of kineticity has a non-negligible impact on the constraints for the other Horndeski parameters due to its effects on the viable parameter space.

The observation of the electromagnetic counterpart to GW170817 has constrained $-6 \times 10^{-15} \leq \alpha^{\rm T}_0 \leq 1.4 \times 10^{-15}$, which seems to eliminate all Horndeski theories with $\alpha_{\rm T} \neq 0$. Typically, the $\alpha_i$'s are parameterized so that they are $0$ in the matter dominated era and only have late time effects \citep[see e.g.][]{linderchallenges,gleyzes2017}. However, the evolution of $\alpha_{\rm T}$ could take a form such that $\alpha_{\rm T} \to 0$ as $z \to 0$, but where $\alpha_{\rm T} \neq 0$ in the past. In this case, \autoref{eq:wave1} is still viable. The power law evolution probed in this paper does not permit this behavior, but other functional forms can. It would be interesting to explore the parameter space and constraints from such a theory.

\acknowledgments
CK would like to thank the referee for their very helpful comments, as well as Marco Raveri, Paul Steinhardt, Anna Ijjas, Alexandre Barreira, Kris Pardo, and Alice Pisani for useful discussions. This work was partially funded through a DAAD (German Academic Exchange Service) Research Grant and the National Science Foundation Graduate Research Fellowship, NSF award number DGE1656466.

\appendix
\section{Model Specifics}
\label{sec:app_bg}

The effective field theory action in unitary gauge and Jordan frame is defined as \citep{cheung,gubitosi,bloomfieldDE}
\begin{align}
\label{eq:action}
S &= \int d^4x\sqrt{-g}\Bigg\{\frac{m_0^2}{2}\left[1+\Omega\left(\tau\right)\right]R+\Lambda\left(\tau\right)-a^2c\left(\tau\right)\delta g^{00} \nonumber \\
&+\frac{M_2^4\left(\tau\right)}{2}\left(a^2\delta g^{00}\right)^2-\frac{\bar{M}_1^3\left(\tau\right)}{2}a^2\delta g^{00}\delta K^\mu{}_\mu \nonumber \\
&-\frac{\bar{M}_2^2\left(\tau\right)}{2}\left(\delta K^\mu{}_\mu\right)^2-\frac{\bar{M}_3^2\left(\tau\right)}{2}\delta K^\mu{}_\nu \delta K^\nu{}_\mu +\frac{a^2\hat{M}^2\left(\tau\right)}{2}\delta g^{00}\delta R^{\left(3\right)} \nonumber \\
&+m^2_2\left(\tau\right)\left(g^{\mu \nu}+n^\mu n^\nu\right)\partial_\mu\left(a^2g^{00}\right)\partial_\nu\left(a^2g^{00}\right)+...\Bigg\} \nonumber \\
&+S_m\left[ \chi_i,g_{\mu \nu}\right]
\end{align}
where
\begin{align}
a^2\delta g^{00}&=a^2g^{00}+1, \\
\delta K^\mu{}_\nu&= K^\mu{}_\nu-K^{\mu 0}{}_\nu, \\
\delta K^\mu{}_\mu&= K^\mu{}_\mu-K^\mu{}_\mu^0, \\
\mbox{and}\quad\delta R^{\left(3\right)}&=R^{\left(3\right)}-R^{\left(3\right)0}
\end{align}
are the perturbations to the time-time metric component, extrinsic curvature, curvature trace, and the three dimensional spatial Ricci scalar, respectively. $n^\mu$~is the normal to surfaces of constant time.  We adopt {\fontfamily{pcr}\selectfont EFTCAMB}'s definitions for clarity.~$\Omega,\,\Lambda,\,\mbox{and}\,c$~are the background evolution equations, as a function of conformal time \citep{EFTCAMB1,EFTCAMB2}. The background equations~$\Lambda\,\mbox{and}\,c$~can be written as a function of~$\Omega$, the coupling to gravity:
\begin{align}
c&=-\frac{m_0^2\ddot{\Omega}}{2a^2}+\frac{m_0^2\mathcal{H}\dot{\Omega}}{a^2}+\frac{m_0^2\left(1+\Omega \right)}{a^2}\left(\mathcal{H}^2-\dot{\mathcal{H}}\right)-\frac{1}{2}\left(\rho_m+P_m\right) \\
\Lambda &=-\frac{m_0^2\ddot{\Omega}}{a^2}-\frac{m_0^2\mathcal{H}\dot{\Omega}}{a^2}-\frac{m_0^2\left(1+\Omega \right)}{a^2}\left(\mathcal{H}^2+2\dot{\mathcal{H}}\right)-P_m,
\end{align}
where~$\rho_m$~and~$P_m$~are the matter energy density and pressure, respectively \citep{EFTCAMB1,EFTCAMB2}. {\fontfamily{pcr}\selectfont EFTCAMB} multiplies~$R$~by~$1+\Omega$~rather than~$\Omega$~for numerical accuracy.~$S_m\left[ \chi_i,g_{\mu \nu}\right]$~is the action for all matter fields~$\chi_i$.

The St{\"u}ckelberg technique makes the scalar perturbations explicit in unitary gauge. The conformal time is perturbed by a scalar field~$\pi$, known as the St{\"u}ckelberg field. All equations are now functions of~$\tau+\pi$, and the perturbation operators transform as \citep{EFTCAMB1,EFTCAMB2}
\begin{align}
\delta g^{00}&\rightarrow \delta g^{00}-2\frac{\dot{\pi}^2}{a^2}-2\frac{\mathcal{H}\pi}{a^2}+... \\
\delta K^\mu{}_\nu&\rightarrow \delta K^\mu{}_\nu + \frac{\dot{\mathcal{H}}}{a}\pi\delta^\mu{}_\nu+\frac{1}{a^2}\bar{\nabla}^\mu\bar{\nabla}_\nu\pi+... \\
\delta R^{\left(3\right)}&\rightarrow \delta R^{\left(3\right)}+4\frac{\mathcal{H}}{a}\bar{\nabla}^2\pi+...
\end{align}

The Horndeski theory of gravity is the most general tensor-scalar theory in which the equations of motion are second order \citep{horndeski}. However, the authors of \cite{zbeyond} and \cite{beyond1} have presented an extended Horndeski theory in which the equations of motion have higher order derivatives. The equations of motion that describe the propagating degrees of freedom reduce to second order equations and, thus, avoid Ostrogradski instabilities \citep[see also][]{beyond2,beyond3}. In this study we restrict ourselves to ordinary Horndeski theory in which operators contain at most two derivatives. The authors of \cite{bloomfield2013} detail the derivatives introduced by the perturbation operators that act on the metric and scalar field perturbations. They note both~$\left(K^\mu{}_\mu\right)^2$~and~$\delta K^\mu{}_\nu\delta K^\nu{}_\mu$~contain terms with four spatial derivatives on the scalar perturbations and with one time and two spatial derivatives. Cancelling the two operators removes the four spatial derivatives, while~$\delta g^{00}\delta R^{\left(3\right)}$ can cancel with the mixed time and spatial derivative term. The~$\partial_\mu\left(a^2g^{00}\right)\partial_\nu\left(a^2g^{00}\right)$~term also contains higher order derivatives that cannot cancel with any other term, so it must be removed. The coefficient relationships required for these cancellations to occur are \citep{bloomfield2013}
\begin{align}
2\hat{M}^2=\bar{M}_2^2=-\bar{M}_3^2\quad\mbox{and}\quad m_2=0.
\end{align}

The authors of \cite{bellini} have formulated a physically motivated parameterization of the coefficients in the action displayed in \autoref{eq:action} for Horndeski theory. The following four parameters are independent of both themselves and the background:
\begin{align}
\label{eq:kineticity}
\alpha_{\rm K} &= \frac{2ca^2+4M_2^4a^2}{m_0^2\mathcal{H}^2\left(1+\Omega+\bar{M}_2^2/m_0^2\right)}, \\
\label{eq:braiding}
\alpha_{\rm B} &= +\frac{a\bar{M}_1^3/m_0^2+a\mathcal{H}\Omega'}{2\mathcal{H}\left(1+\Omega+\bar{M}_2^2/m_0^2\right)}, \\
\label{eq:tensor}
\alpha_{\rm T} &= -\frac{\bar{M}_2^2}{m_0^2\left(1+\Omega\right)+\bar{M}_2^2}, \\
\label{eq:alpha_M}
\mathrm{and}\quad \alpha_{\rm M} &= a\frac{d \ln M_*^2}{d a}=\frac{a\left(\Omega+\bar{M}_2^2/m_0^2\right)'}{1+\Omega+\bar{M}_2^2/m_0^2},
\end{align}
where primes denote derivatives with respect to the scale factor~$a$. In our work we choose a $\Lambda \mbox{CDM}$ background. {\fontfamily{pcr}\selectfont EFTCAMB} evolves~$M^2_*/m^2_0=1+\widetilde{M}=1+\Omega+\bar{M}_2^2 / m_0^2$~for numerical reasons, transforming \autoref{eq:alpha_M} to $\alpha_{\rm M}=a\widetilde{M}'/\left(1+\widetilde{M}\right)$.

\section{Effects of Kineticity on Stability Constraints}
\label{sec:stability_constraints}

Kineticity's effects on the observables are closely related to the accuracy of the quasi-static approximation. The authors of \cite{sawicki2015} have shown that when analyzing the CMB, the quasi-static limit should be used neither when dark energy has a non-negligible effect at recombination nor when modeling the integrated Sachs-Wolfe (ISW) effect. If the dark energy sound speed, also known as the $\pi$ field sound speed, is less than 0.1, the quasi-static limit is not valid for CMB lensing. When the approximation is valid, kineticity does not enter the equations of motion and is not well constrained by observations. As mentioned in Section \ref{sec:data}, Ref. \cite{bellini2015} found that $\alpha_{\rm K}$ could not be well constrained with their parameterization. They then presented constraints on $\alpha_{\rm T},\,\alpha_{\rm M},\,$and$\,\alpha_{\rm B}$ for a few fixed values of kineticity. To limit the number of additional degrees of freedom in our analysis, we also fix $\alpha_0^{\rm K}$ in our analysis.

With $\alpha_0^{\rm T}=0$, we find that evolving the remaining parameters as constants, with $\alpha_{\rm K} = 0.1$, yields a stable parameter space too small to explore with an MCMC analysis. Evolving the Horndeski parameters as power laws (see Section \ref{sec:data} and Appendix \ref{sec:evolution}) enlarges the stable parameter space and provides the opportunity to probe the time evolution of the parameters.

\begin{figure*}[h]
\begin{center}
\includegraphics[width=\textwidth]{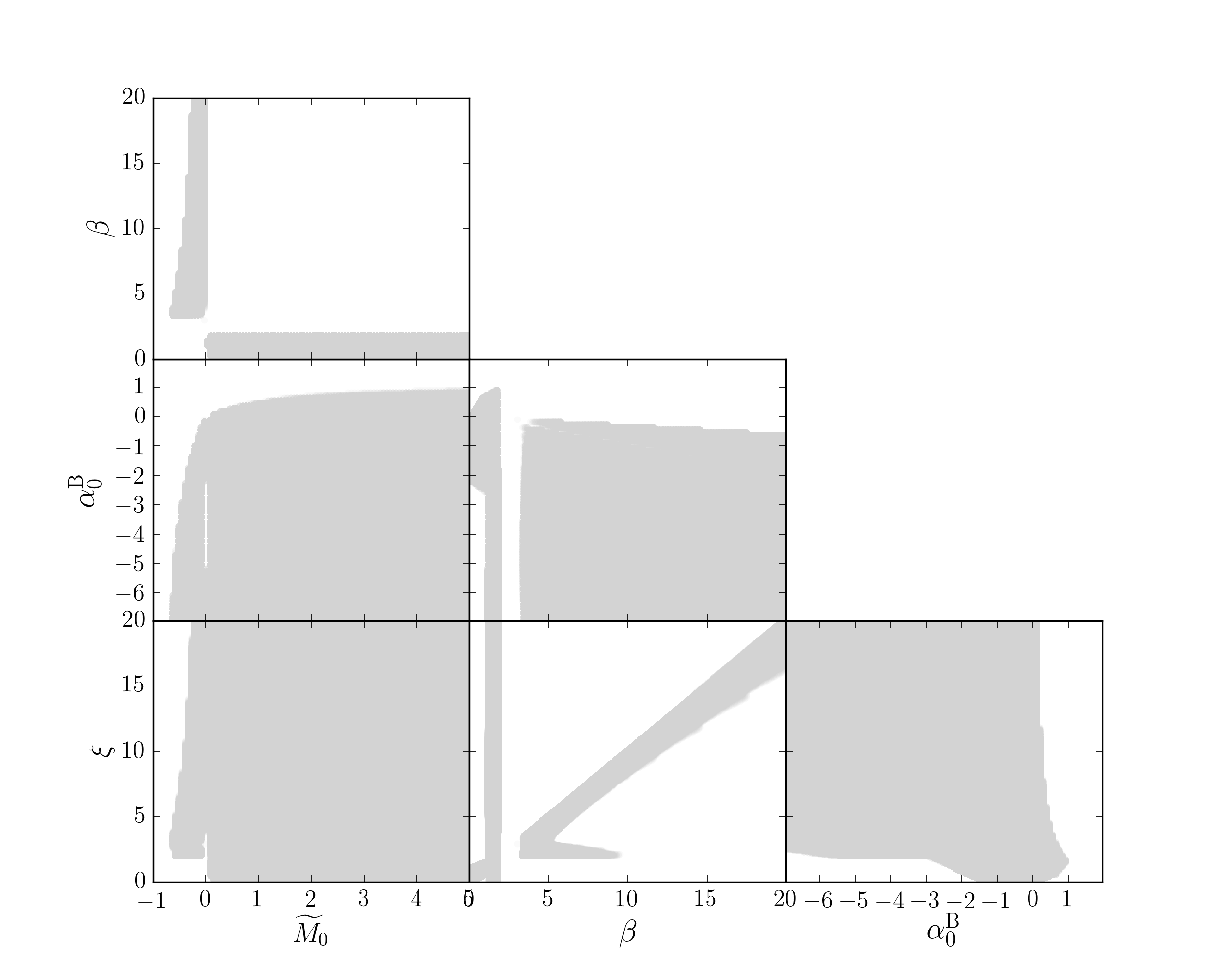}
\caption{Stable parameter space when $\alpha_{\rm K}=0.1a^3$.}
\label{fig:stab_Kpt1}
\end{center}
\end{figure*}

\begin{figure*}[h]
\begin{center}
\includegraphics[width=\textwidth]{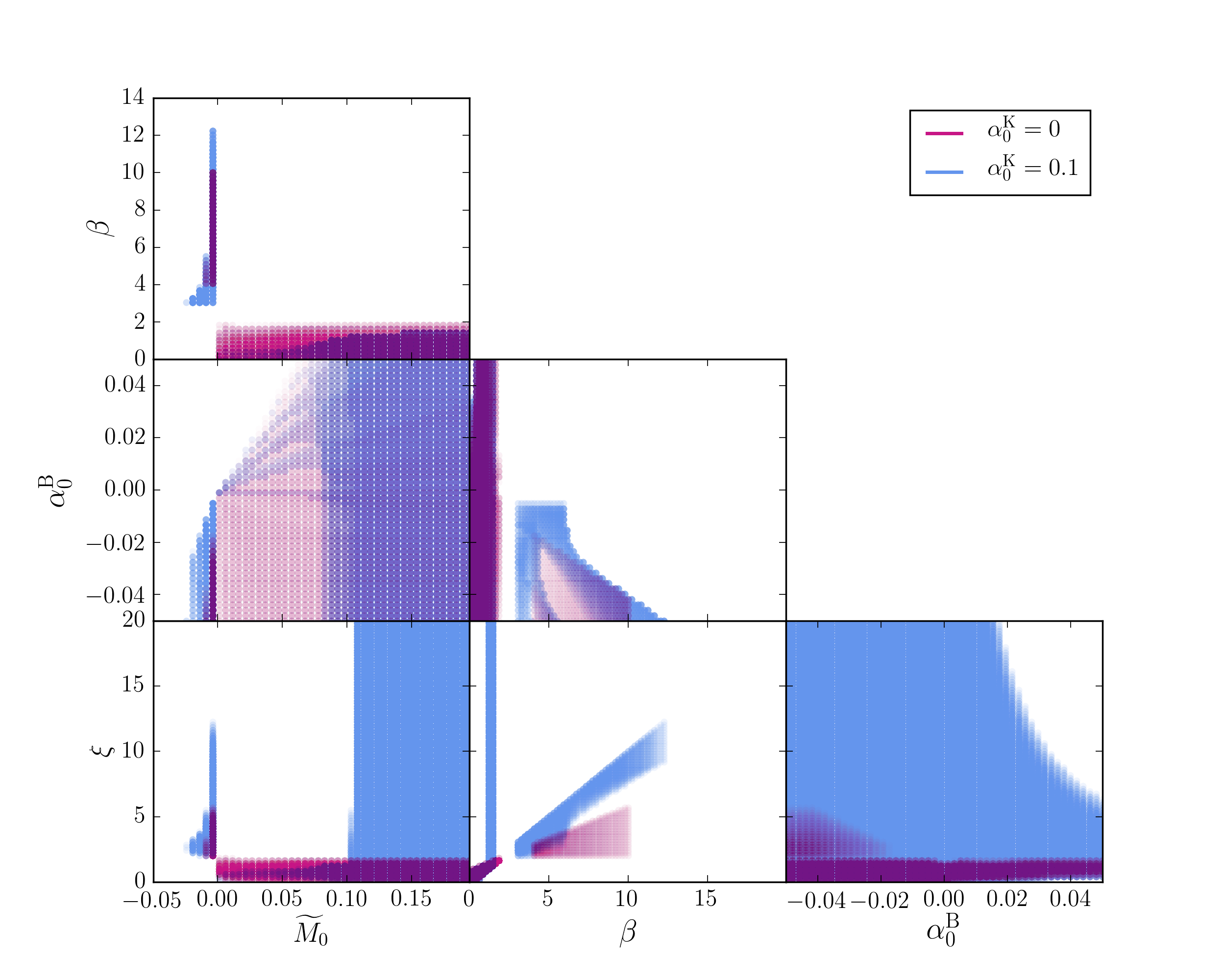}
\caption{Comparison between the stable parameter spaces for $\alpha_0^{\rm K}=0$ and $\alpha_0^{\rm K}=0.1$ when $\alpha_0^{\rm B}$ is near $0$.}
\label{fig:stab_Kpt1_K0}
\end{center}
\end{figure*}

Although the authors of \cite{bellini2015} found the remaining Horndeski parameters have a weak dependence on kineticity, we note that fixing the kineticity has non-trivial effects on our viable parameter space. {We believe this is primarily due to the tachyon instabilities defined in Section \ref{sec:stability}.} In \autoref{fig:stab_Kpt1} we explore the stable parameter space for $\alpha_0^{\rm K} = 0.1$. We fix $\kappa=3.0$, which we found provides a large range of likely values for the other Horndeski parameters. The stable region for $\beta$, the power law index related to friction, has a gap from $2 \lesssim \beta \lesssim 3$ due to mathematical stability conditions. Without using a nested sampling method, the MCMC cannot reach both stable $\beta$ regions. We choose to only explore the smaller friction exponent values $\beta \lesssim 2$ since large values will drive the friction parameter $\alpha_{\rm M}$ close to $0$. Having $\widetilde{M}_0 \approx 0$ effectively achieves the same result if the data prefer $\alpha_{\rm M} \approx 0$.

Stability requires the scalar propagation speed to satisfy \citep{bellini}
\begin{align}
  c_s^2=-\frac{2\left(1+\alpha_{\rm B}\right)\left[\dot{\mathcal{H}}-\left(1-\alpha_{\rm B}+\alpha_{\rm M}\right)\mathcal{H}^2\right]+2\mathcal{H}\dot{\alpha}_{\rm B}+a^2\left(\tilde{\rho}_{\rm m}+\tilde{p}_{\rm m}\right)}{H^2\left(\alpha_{\rm K}+6\alpha_{\rm B}^2\right)} > 0,
\end{align}
using the definition for $\alpha_{\rm B}$ of \cite{eftcambnotes}. Because both kineticity and braiding are in the denominator, it is not stable to have $\alpha_0^{\rm K}=0$ and $\alpha_0^{\rm B}=0$ simultaneously. Thus, to explore the $\alpha_0^{\rm B}=0$ case and compare with the $\alpha_0^{\rm B} \neq 0$ case in a self-consistent manner, we must fix $\alpha_0^{\rm K}$ at a nonzero value.

We analyzed the stable parameter space for $\alpha_0^{\rm K} = 0$ and $\alpha_0^{\rm K} = 1.0$, as well. Increasing $\alpha_0^{\rm K}$ from $0.1$ to $1.0$ significantly shrinks the stable parameter space to the extent it cannot be explored with an MCMC analysis. We fix $\alpha_0^{\rm K}=0.1$ throughout our analysis due to its larger stable parameter space and closeness to GR.

In \autoref{fig:stab_Kpt1_K0} we compare the stable parameter space for $\alpha_0^{\rm K}=0$ and $\alpha_0^{\rm K}=0.1$ for values near $\alpha_0^{\rm B}=0$. The higher the point opacity the larger the stable parameter space allowed by the parameter values describing that point. Regions of overlap indicate where parameters are stable for both kineticity values. The difference in the stable parameter space for the two kineticity values is dramatic, indicating the importance of understanding how all imposed stability conditions affect the viable parameter space to correctly interpret the parameter posteriors. When $\alpha_0^{\rm K}=0.1$, the viable parameter space when $0 \lesssim \widetilde{M}_0 \lesssim 0.1$ is small, whereas for $\alpha_0^{\rm K}=0$ the same $\widetilde{M}_0$ values provide a larger stable parameter space. {The tachyon stability conditions solely drive these differences and those in the other stability contours.} For the $\alpha_0^{\rm K}=0.1$ case, the viable $\widetilde{M}_0-\alpha_0^{\rm B}$ space converges to a single point as both $\alpha_0^{\rm B}$ and $\widetilde{M}_0$ approach $0^+$, i.e. their GR value from the right. Thus, the choice of kineticity affects the $\widetilde{M}_0$ posterior {due to the tachyon stability constraints} (see Section \ref{sec:results}). Indeed, if we remove the mathematical stability conditions the viable parameter spaces for $\alpha_0^{\rm K}=0$ and $\alpha_0^{\rm K}=0.1$ are identical.

We also note that while the stability contours for $\alpha_0^{\rm K}=0$ make it appear $\alpha_0^{\rm B}=0$ is stable for this kineticity, there are no stable points when both $\alpha_0^{\rm K}=0$ and $\alpha_0^{\rm B}=0$ \textit{exactly}. The appearance of the plot is an artifact of discretely sampling the parameter space.

\section{Parameter Evolution}
\label{sec:evolution}

\begin{figure}[h]
\begin{center}
\includegraphics[width=0.67\textwidth]{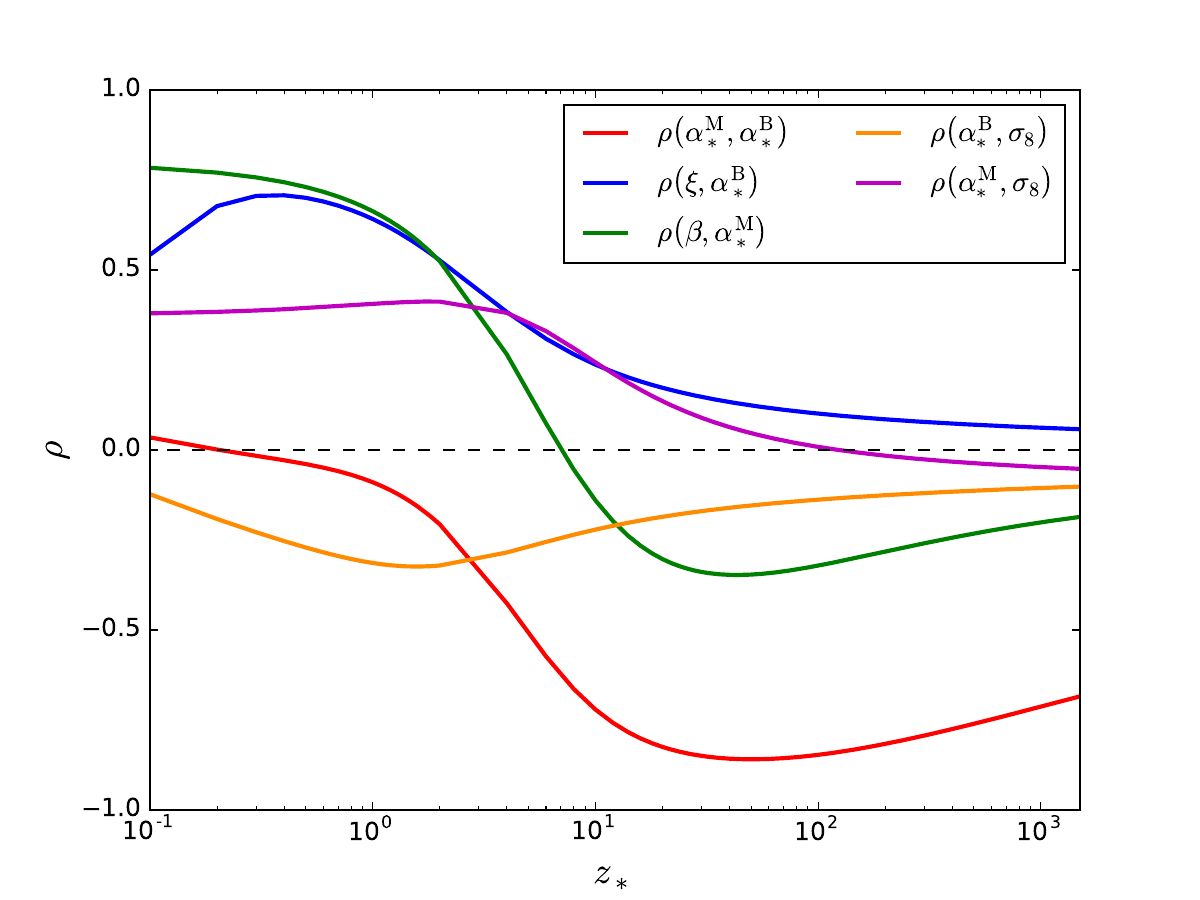}
\caption{Correlation coefficients for the Horndeski parameters derived from the CMB+LSS chains.}
\label{fig:corr_coeff}
\end{center}
\end{figure}

To better constrain the Horndeski parameters we explore the pivot point $a_{*}$ at which to measure the parameters. The parameters of interest to measure then become
\begin{align}
\alpha^{\rm B}_{*}&=\alpha^{\rm B}_0a^{\xi}_{*} \\
\alpha^{\rm M}_{*}&=\widetilde{M}_0a_{*}^{\beta}\frac{\beta}{1+\widetilde{M}_0a_{*}^{\beta}}
\end{align}
along with their exponents $\xi$ and $\beta$. We are then evolving
\begin{align}
\alpha_{\rm B}&=\alpha^{\rm B}_{*}\left(\frac{a}{a_{*}}\right)^\xi \\
\alpha_{\rm M}&=\alpha^{\rm M}_*\left(\frac{a}{a_*}\right)^\beta\frac{\beta}{\beta+\left[\left(a/a_*\right)^\beta-1\right]\alpha^{\rm M}_*}.
\end{align}
We can derive the braiding and friction today, $\alpha^{\rm B}_0$ and $\widetilde{M}_0\beta/(1+\widetilde{M}_0)$, respectively, from these measured values. In Figure \ref{fig:corr_coeff} we show the correlation coefficient between Horndeski parameters of interest as well as $\sigma_8$ as a function of pivot redshift $z_*$. $\alpha_*^{\rm B}$ is relatively uncorrelated with the other parameters at decoupling, which could be an artifact of the power law evolution. $\alpha_*^{\rm M}$ and $\sigma_8$ are positively correlated at present. This is consistent with the effects seen in the matter power spectrum when increasing $\widetilde{M}_0$ (see Section \ref{sec:constrain}). Large scales entered the horizon at low redshifts when $\alpha_*^{\rm M}$ and $\sigma_8$ were positively correlated. Hence, increasing $\widetilde{M}_0$ creates an amplifying effect on large scales. $\alpha_*^{\rm B}$ and $\alpha_*^{\rm M}$ are positively correlated with their respective exponents at the present epoch. However, the only epoch during which $\alpha_*^{\rm B}$ and $\alpha_*^{\rm M}$ are uncorrelated is at present.

$\alpha^{\rm B}_{*}$ has the smallest relative uncertainty at present. The relative uncertainty for $\alpha^{\rm M}_{*}$ has a small minimum near $z_* \approx 10-20$, similar redshifts to where $\alpha^{\rm M}_{*}$ is uncorrelated with $\beta$ and $\sigma_8$. This suggests a pivot point earlier than the present may be preferable for $\alpha_{\rm M}$. However, the relative uncertainty for $\alpha^{\rm M}_{*}$ at present is almost as small. Thus, because $\alpha_*^{\rm B}$ and $\alpha_*^{\rm M}$ are uncorrelated at present and the relative uncertainties for each are minimized or close to the minimum, we choose to keep the pivot point at present.

\vfill
\eject
\bibliography{draft}
\end{document}